\documentclass[aps,floatfix,showpacs,nofootinbib,showkeys,twocolumn]{revtex4}
\usepackage{amsmath,amsfonts,euscript,amssymb}
\usepackage{colordvi}
\usepackage[T1]{fontenc}
\usepackage[english]{babel}
\usepackage{epsfig}
\usepackage{graphicx}

\newcommand{\be}{\begin{equation}}
\newcommand{\ee}{\end{equation}}
\newcommand{\ba}{\begin{eqnarray}}
\newcommand{\ea}{\end{eqnarray}}
\newcommand\nn{\nonumber}
\newcommand{\partialslash}{\partial\hspace{-.5em}/\hspace{.15em}}

\begin{document}

\title{Two-photon decays and photoproduction on electrons of $\eta(550)$,
       $\eta'(958)$, $\eta(1295)$, and $\eta(1475)$ mesons}

\author{A.\ B.\ Arbuzov}
\affiliation{Bogoliubov Laboratory of Theoretical Physics,
JINR, Dubna, 141980  Russia}
\affiliation{Department of Higher Mathematics, University Dubna,
Dubna, 141980   Russia}

\author{M.\ K.\ Volkov}
\affiliation{Bogoliubov Laboratory of Theoretical Physics,
JINR, Dubna, 141980  Russia}


\begin{abstract}
Electromagnetic interactions of the ground and first radial excited states of $\eta$ and $\eta'$
mesons in the framework of the extended $U(3)\times U(3)$ NJL model are considered.
The radial excitations are described with the help of polynomial form factor of
the second order over the inner quark momentum.
The solution of the $U_A(1)$ problem by means of 't~Hooft interaction is taken into account.
For diagonalization of the free Lagrangian the $4\times4$ mixing matrix is used.
Two-photon decay widths of the ground $\eta$ and $\eta'$ meson are found to be in
a satisfactory agreement with the experiment. Predictions for the two-photon decay widths of
$\eta(1295)$ and $\eta(1475)$ are given. The probabilities of eta meson production by two-photon
mechanism in $e^+e^-$ collisions and of their photoproduction on electrons are calculated.
\end{abstract}

\date{\today}



\maketitle

\section{Introduction}

At the present time extensive investigations of meson production
in electron-positron collisions of different energies
are carried on at various experimental facilities like
VEPP-2000 (Novosibirsk), DA$\Phi$NE (Frascati), BEPC-II (Beijing), KEK-B (KEK)
and other. For theoretical description of these processes at low energies
one can not use the perturbative QCD, therefore it is necessary to use
phenomenological models: the vector meson dominance one,
the ones based on the chiral symmetry {\it etc.}

In a set of recent papers
we used the extended $SU(2)\times SU(2)$ Nambu--Jona-Lasinio (NJL) model
with a polynomial form factor~\cite{Volkov:1996br,Volkov:1997dd} for description of
electromagnetic interactions of $\pi$, $\rho$, and $\omega$ mesons
and of their first radial excited states.
In Ref.~\cite{Kuraev:2009uh} the two-photon decays of $\pi$ and $\pi'\equiv\pi(1300)$
mesons and the processes of their production in $e^+e^-$ collisions were considered.
In papers~\cite{Arbuzov:2010xi,Arbuzov:2011fv} we computed the production cross sections of
$\pi^0(\pi')\omega$ and $\pi^0(\pi')\gamma$ pairs at electron positron colliders.
These results were in a satisfactory agreement as with the experimental
data~\cite{Akhmetshin:2003ag,Achasov:2000wy,Achasov:2000zd,Achasov:2003ed}
as well as with theoretical estimates~\cite{Akhmetshin:2003ag}
obtained within the vector meson dominance model.
Radiative decays of $\pi^0$, $\rho^0$, $\omega$, and their radial excited states
were studied in Ref.~\cite{Arbuzov:2010vq}.

In this work we describe the two-photon decays of pseudoscalar isoscalar $\eta$, $\eta'(958)$,
$\eta(1295)$, and $\eta(1475)$ mesons, the processes of their production in two-photon
mechanism at electron-positron collisions, and in the Primakoff effect on an electron.
The calculations are performed on the base of the $U(3)\times U(3)$ chiral NJL model
with 't~Hooft interaction. The radial excitations of mesons are described again with the
help of polynomial form factors for meson interactions with $u(d)$ and $s$
quarks~\cite{Volkov:1996fk}. Diagonalization of the free Lagrangian for the four eta-mesons
with the transition to their physical states is performed by means of the $4\times 4$ mixing
matrix~\cite{Volkov:1999iq,Volkov:1999xf,Volkov:1999yi}. Note that we skip the $\eta(1405)$ state
which is usually treated as a pseudoscalar glueball~\cite{Crede:2008vw,Mathieu:2011mq,Cheng:2008ss}.

\section{Lagrangian}

We use a  nonlocal separable four-quark interaction
of a current-current form which admits nonlocal vertexes (form
factors) in the quark currents, and a pure local six-quark 't Hooft
interaction~\cite{Vogl:1991qt,Klevansky:1992qe}:
\ba
&&     {\cal L}(\bar q, q) =
     \int\! d^4x\; \bar q(x)
     (i \partialslash -m^0) q(x)+
     {\cal L}^{(4)}_{\rm int}+
     {\cal L}^{(6)}_{\rm int},  \label{lag}
     \nonumber \\ \nn
&&     {\cal L}^{(4)}_{\rm int} =
     \int\! d^4x\sum^{8}_{a=0}\sum^{N}_{i=1}
     \frac{G}{2}[j_{S,i}^a(x) j_{S,i}^a(x)+
     j_{P,i}^a(x) j_{P,i}^a(x)],\\
&&     {\cal L}^{(6)}_{\rm int}=-K \left[\det
     \left[\bar q (1+\gamma_5)q\right]+
     \det\left[\bar q (1-\gamma_5)q\right]
     \right].
\ea
Here, $m^0$ is the current quark mass matrix ($m_u^0\approx m_d^0$) and
$j^a_{S(P),i}$ denotes the scalar (pseudoscalar) quark currents
\be
     j^a_{S(P),i}(x)=
     \int\! d^4x_1 d^4x_2\; \bar q(x_1)
     F^a_{S(P),i }(x;x_1,x_2) q(x_2)
\ee
where $ F^a_{S(P),i}(x;x_1,x_2)$ are the scalar (pseudoscalar)
nonlocal quark vertex. The coupling constants $G=3.14$~GeV$^{-2}$
and $K=6.1$~GeV$^{-5}$ are fixed in the model~\cite{Volkov:1999xf,Volkov:1999yi} from the pion mass and
from the masses of $\eta$ and $\eta'$ mesons (taking into account the mixing of ground and
excited states), respectively.

To describe the first radial excitations of mesons, we take
the form factors in momentum space as follows (see \cite{Volkov:1996br,Volkov:1997dd}):
\ba
&& F_{P,j}^a= i\gamma_5 \lambda^a f^a_j,
\nonumber \\
&&    f^a_1\equiv 1,\quad f^a_2\equiv f_a({\bf k})=c_a(1+d_a {\bf k}^2),\label{fDef}
\ea
where
$\lambda^a$ are Gell--Mann matrices,
$\lambda^0 = \sqrt{\frac{2}{3}}\cdot${\bf 1}, with {\bf 1} being the unit matrix.
Here, we consider the form factors in
the rest frame of mesons
\footnote{The form factors depend on the transverse parts of
the relative momentum of quark-antiquark pairs $k_{\perp} =
k - \frac{k\cdot P}{P^2}P$, where $k$ and
$P$ are the relative and total momenta of a quark-antiquark pair,
respectively. Then, in the rest frame of mesons, ${\bf P}_{\mathrm{meson}}$ = 0,
the transverse momentum is
$k_{\perp } = (0, {\vec k})$,
and we can define the form factors as depending on the 3-dimensional momentum ${\vec k}$ alone.
}%
. The slope parameters $d_u=-1.78$~GeV$^{-2}$ and  $d_s=-1.73$~GeV$^{-2}$) are defined from the
condition that the tadpoles with the form factors for the corresponding quarks are equal to zero.
That means that the excited states do not give contributions to the quark condensates.
The coefficients $c_u=1.5$, $c_s=1.66$ are fitted from the masses of the radial excited physical states
of eta-mesons.

Following~\cite{Vogl:1991qt,Klevansky:1992qe},
after coupling of a pair of quarks in the 't~Hooft interaction we get a modified four-quark
interaction of the isoscalar pseudoscalar sector of the NJL model:
\ba \label{q-meson}
     {\cal L}_{\rm isosc}=\sum_{a,b=8}^9
     (\bar q i\gamma_5\tau_a q)T^P_{a b}
     (\bar q i\gamma_5\tau_b q),
\ea
where $T^{P}$ is a matrix with elements defined as follows
\ba
&&     T^{P}_{88}=G^{+}_{u}/2,\quad  T^{P}_{89}=G^{+}_{us}/2,
\nn \\
&&     T^{P}_{98}=G^{+}_{us}/2,\quad  T^{P}_{99}=G^{+}_{s}/2,
\nn \\
&&   \tau_8 = \lambda_u = ({\sqrt 2}
     \lambda_0 + \lambda_8)/{\sqrt 3},
\nn \\
&&   \tau_9 = \lambda_s = (-\lambda_0 +
     {\sqrt 2}\lambda_8)/{\sqrt 3},
\nn \\
&&   G_u^{+}= G - 4Km_sI_1(m_s), \qquad
     G_s^{+}= G,
\nn \\
&&   G_{us}^{+}=  4{\sqrt 2}Km_uI_1(m_u).\nonumber
\ea
Here $m_u$ and $m_s$ are the constituent quark masses and $I_1(m_q)$
is the integral which for an arbitrary $n$ is defined as follows
\be
     I_n^{f\ldots f_a}(m_a)={-i \frac{N_c}{(2\pi)^4}}
     \int_{\Lambda_3}\!d^4 k
     \frac{f_a({\bf k})\ldots f_a({\bf k})}{(m^2_a-k^2)^n} .
     \label{DefI}
\ee
The 3-dimensional cut-off $\Lambda_3=1.03$~GeV in  is implemented
to regularize the divergent integrals.
Here $m_{u,s}$ are the constituent quark masses: $m_u=280$~MeV and $m_s=405$~MeV.

After bosonization we get the quark-meson interaction Lagrangian in the form
\ba\label{L_int}
&& {\cal L}_{\mathrm{int}} = \bar{q}(p_1)\biggl[
(i\slash{\!\!\! p}-m)
+ i\gamma_5 \biggl(
\lambda_u g_{1}^{u} \phi_1^{u}
+ \lambda_s g_{1}^{s} \phi_1^{s}
\nonumber \\
&& \qquad +\lambda_u g_{2}^{u} \phi_2^{u} f_2^u({\bf p})
+ \lambda_s g_{2}^{s} \phi_2^{s} f_2^s({\bf p})
\biggr)
\biggr]q(p_2),
\ea
where $m=diag(m_u,m_d,m_s)$, $p=p_1-p_2$,
\ba
&& g_1^u = [Z4I_2^u]^{-1/2}, \qquad
g_1^s = [Z4I_2^s]^{-1/2},
\nonumber \\
&& g^{u,s}_2=[4I_2^{ff_{u,s}}(m_{u,s})]^{-1/2},
\ea
where $\pi-a_1$ transitions accounted by factor
\ba
Z = 1 - \frac{6m_u^2}{M_{a_1}^2}
\ea
taken the same for ground $\phi_u$ and $\phi_s$ meson states. For the excited
states these transitions can be omitted as discussed in Ref.~\cite{Volkov:1996fk}.

Note that here we do not take into account transitions between
pseudoscalar and axial-vector states, which were important for
description of pions and kaons. This leads to a certain change
with respect to the previous works~\cite{Volkov:1999iq,Volkov:1999xf,Volkov:1999yi}.

From Eq.~(\ref{q-meson}) taking into account renormalization of the kinetic terms
in the one-loop approximation,
one can get the free meson Lagrangian in the following form~\cite{Volkov:1999iq,Volkov:1999xf}:
\ba
&& {\cal L}^{(2)}(\phi)=\frac12\sum_{i,j=1}^2\sum_{a,b=8}^{9}
          \phi_i {\cal K}_{\phi,ij}(P)\phi_j,
    \label{L2a} \\ \nonumber
&& \phi=(\phi_1^8,\phi_2^8,\phi_1^9,\phi_2^9),
\ea
where
\ba
     {\cal K}_{\phi,11}(P)&=&P^2-(m_u\pm m_u)^2-M_{\phi^8,1}^2,\nonumber\\
     {\cal K}_{\phi,22}(P)&=&P^2-(m_u\pm m_u)^2-M_{\phi^8,2}^2,\nonumber\\
     {\cal K}_{\phi,33}(P)&=&P^2-(m_s\pm m_s)^2-M_{\phi^9,1}^2,\nonumber\\
     {\cal K}_{\phi,44}(P)&=&P^2-(m_s\pm m_s)^2-M_{\phi^9,2}^2,\\
     {\cal K}_{\phi,12}(P)&=&{\cal K}_{\phi,21}(P)=\Gamma_{\eta_u}(P^2-(m_u\pm m_u)^2),\nonumber\\
     {\cal K}_{\phi,34}(P)&=&{\cal K}_{\phi,43}(P)=\Gamma_{\eta_s}(P^2-(m_s\pm m_s)^2),\nonumber\\
     {\cal K}_{\phi,13}(P)&=&{\cal K}_{\phi,31}(P)=
	   g_1^u g_1^s (T^{P})^{-1}_{89},
\nonumber \\
\Gamma_{\eta_{u,s}} &=& \frac{I_2^{f_{u,s}}(m_{u,s})\sqrt{Z}}{\sqrt{I_2(m_{u,s}) I_2^{ff_{u,s}}(m_{u,s})}}
. \nonumber
\ea
The {\em bare} meson masses are
\ba
     && M_{\phi^8,1}^2= (g_1^{u})^{2}
	\left({1\over 2}(T^{P})^{-1}_{88} - 8I_1(m_u) \right),
     \nonumber \\
     && M_{\phi^9,1}^2= (g_1^{s})^{2}
	\left({1\over 2}(T^{P})^{-1}_{99}-8I_1(m_s) \right), \nonumber\\
     && M_{\phi^8,2}^2=(g_2^{u})^{2}
	\left({1\over 2G} - 8I_1^{ff_u}(m_u) \right),\\
     && M_{\phi^9,2}^2=(g_2^{s})^{2}
	\left({1\over 2G}-8I_1^{ff_s}(m_s) \right). \nonumber
     \label{Mpuu}
\ea
Transition from the {\em bare} states to the {\em physical} ones is performed with the
help of $4\times4$ matrix $R$ which provides the diagonal form for the free Lagrangian.
This matrix was found numerically in~\cite{Volkov:1999iq,Volkov:1999xf}, 
it is given in Table~\ref{table:1}.
Note that this matrix is not unitary. This is due to the fact that
it contains not only a rotation but also dilations.
The dilations are related to two different renormalizations: one of them for
the ground states and one for the first radial excited states.
One can observe the same feature of non-unitary transformation in a more
simple case SU(2)$\times$SU(2) extended NJL model~\cite{Volkov:1996br}
where the mixing of $\pi$ and $\pi'$ mesons was considered.
It is worth to note that if we exclude from the consideration the excited states,
we would get instead of the $4\times4$ matrix a simple $2\times2$ matrix corresponding
to unitary orthogonal transformations with the standard singlet-octet mixing angle
$\theta\approx -19^{\circ}$, see Ref.~\cite{Volkov:1999qb}. The latter quantity
is close to the values of the mixing angle obtained in numerous theoretical
and experimental studies, see {\it e.g.} paper~\cite{Pham:2010sr} and references therein.  

\begin{table}
\caption{The mixing coefficients for the isoscalar pseudoscalar meson states.}
\label{table:1}
$$
	\begin{array}{|r|c|c|c|c|}
	\hline
	 R_{i,j}		&\eta 		&\hat\eta 	&\eta' 		&\hat\eta'\\
	\hline
	\varphi^8_{1}	&0.71		&0.62		&-0.32		&0.56		\\
	\varphi^8_{2}   &0.11		&-0.87		&-0.48		&-0.54		\\
	\varphi^9_{1}	&0.62		&0.19		&0.56		&-0.67		\\
	\varphi^9_{2}   &0.06		&-0.66		&0.30		&0.82		\\
	\hline
	\end{array}
$$
\end{table}
%

\section{Description of radiative processes}

\subsection{Two-photon decays}

Let us start with the 2-photon decays of the ground and excited states of $\eta$-mesons.
To describe it we introduce the quark-photon interaction term
$\bar{q}Q\gamma^{\mu}A^{\mu}q$ into the interaction Lagrangian~(\ref{L_int}),
with $Q=diag(2/3,-1/3,-1/3)$. For the decay $\eta\to\gamma\gamma$ we get the
amplitude
\ba
&& T_{\eta\to\gamma\gamma} = \frac{\alpha}{9\pi}
\varepsilon_{\mu\nu\rho\sigma}q_1^{\rho}q_2^{\sigma}\varepsilon_1^\mu\varepsilon_2^\nu
\nonumber \\ && \quad
\biggl\{
 R_{1,1} \frac{g_1^{u}}{m_u} 5 I_3(m_u)
+R_{2,1} \frac{g_2^{u}}{m_u} 5 I_3^{f_u}(m_u)
\nonumber \\ && \quad
-R_{3,1} \frac{g_1^{s}}{m_s} \sqrt{2} I_3(m_s)
-R_{4,1} \frac{g_2^{s}}{m_s} \sqrt{2} I_3^{f_s}(m_s)
\biggr\}.
\ea
Analogously we got the amplitudes for two-photon decays of $\hat\eta$, 	$\eta'$, and $\hat\eta'$
mesons. Our results for the widths of the four decays with comparison to the existing
experimental data are given in Table~\ref{table:2}.

\begin{table}
\caption{Widths of $\eta$ meson two-photon decays.}
\label{table:2}
$$
	\begin{array}{|c|c|c|c|c|}
	\hline
	 \mathrm{meson}		 &\eta    & \hat\eta 	 & \eta' 	& \hat\eta' \\
	\hline
     \mathrm{model~[eV]}  & 520      & 93        & 4990         &  230 \\
    \hline
     \mathrm{exp.~[eV]}  & 510\pm26  & -       & 4340\pm140    &  - \\
	\hline
	\end{array}
$$
\end{table}
%

Consider now the processes of $\eta$-meson production in $e^+e^-$ collisions
by two-photon mechanism:
\ba
e^+ + e^- \to e^+ + e^- + \eta(\hat\eta,\;\eta'\;\hat\eta').
\ea
In the first approximation the total cross section~\cite{Brodsky:1970vk} reads
\ba
&& \sigma_{\eta_i}= (4\alpha)^2 \ln^2\frac{\sqrt{s}}{2m_e}
\frac{\Gamma(\eta_i\to2\gamma)}{M^3(\eta_i)}
Y\biggl(\frac{M^2(\eta_i)}{s}\biggr),
\nonumber \\
&& Y(z)=(2+z)^2\ln\frac{1}{\sqrt{z}}-(3+z)(1-z).
\ea
The corresponding numerical results are given in Table~\ref{table:3} in comparison
with the experimental data~\cite{Roe:1989qy} existing for the ground meson states
at $\sqrt{s}=29$~GeV total $e^+e^-$ energy in the center-of-mass.

\begin{table}
\caption{Total cross sections of $\eta$ meson production via the two-photon mechanism.}
\label{table:3}
$$
	\begin{array}{|c|c|c|c|c|}
	\hline
	 \mathrm{meson}		 &\eta           & \hat\eta 	& \eta' 	 & \hat\eta' \\
	\hline
     \mathrm{model~[nb]} & 1.4           &  0.014       & 2.1       &  0.022 \\
    \hline
     \mathrm{exp.~[nb]}  & 1.25\pm 0.13  & -            & 1.8\pm0.3 &  - \\
	\hline
	\end{array}
$$
\end{table}
%

In a similar manner we can describe the Primakoff process of $\eta$ meson production
in photon--lepton collisions
\ba
&& \gamma(k) + l(p) \to \eta_i(p_1) + l(p'), \qquad l = e,\mu,
\nn \\
&& p^2={p'}^2=m_l^2, \quad k^2 = 0, \quad p_1^2=M^2(\eta_i),
\nn \\
&& s = 2kp > M^2(\eta_i) \gg m_l^2.
\ea
The total cross section of this process reads~\cite{Kuraev:2009uh}:
\ba
&& \sigma^{\gamma l\to\eta_i l} = \frac{\alpha\Gamma}{M^3(\eta_i)}
\biggl[1+\biggl(1-\frac{M^2(\eta_i)}{s}\biggr)^2\biggr]
\nn \\ && \quad \times
\biggl( \ln\frac{s^2}{m_l^2M^2(\eta_i)}-1\biggr).
\ea
For electron-photon collisions at the center-of-mass energy $\sqrt{s}=3$~GeV
we get the following predictions for the total cross sections:
$\sigma^{\gamma e\to\eta e} = 340$~pb,
$\sigma^{\gamma e\to\hat\eta e} = 540$~pb,
$\sigma^{\gamma e\to\eta' e} = 3.7$~pb,
$\sigma^{\gamma e\to\hat{\eta'} e} = 5.7$~pb.

\section{Conclusions}

The results of our calculations show that the application of the mixing matrix $R$
which diagonalizes the free meson Lagrangian leads to sufficiently good description
radiative decays of the ground $\eta$ and $\eta'$ meson states. This allows us to hope
to get reasonable predictions for their first radial excited states. A similar situation took place
in the case of strong interactions of the ground and excited mesons~\cite{Volkov:1997dd,Volkov:1999yi}.

The $\eta(1475)$ meson as a $q\bar{q}$ state was considered in Ref.~\cite{Achasov:2011xc}, where
a simple phenomenological model of mixing between only two $u\bar{u}+d\bar{d}$ and $s\bar{s}$ components
was used. In our approach we have mixing of four $q\bar{q}$ components: two of them for ground states
and two for radial excited ones. However we did not take into account mixing of $\eta$ mesons with
pseudoscalar glueballs in particular with the $\eta(1405)$ state.
Meanwhile we took into account the influence of the gluon anomaly
(by means of the 't~Hooft interaction) onto the mixing of the
$u\bar{u}+d\bar{d}$ and $s\bar{s}$ components in all four states being considered.
Some other papers were devoted to studies of mixing between the ground $\eta(550)$ and $\eta'(958)$
mesons with a pseudoscalar glueball, see {\it e.g.} Refs.~\cite{Cheng:2008ss,Anisovich:1997dz,Kiesewetter:2010ze}.
Note that the excited $\eta$ meson states were not taken into account there. Moreover, the mixing of the ground
$\eta$ and $\eta'$ mesons with glueball was found to be not large. 
We argue that in future studies all these mixing effects should be considered together like it
was done in ref.~\cite{Volkov:2001ct} for mixing of four isoscalar mesons and a glueball using $5\times5$
mixing matrix.


\subsection*{Acknowledgments}
We are grateful to E.A.~Kuraev for fruitful discussions and critical reading of the manuscript.
This work was supported by RFBR grant 10-02-01295-a.

\end{document}